\documentclass[journal]{IEEEtran}
\ifCLASSINFOpdf
\else
\fi

\usepackage{multirow}
\usepackage{cite}
\usepackage{amsmath,amssymb,amsfonts}

\DeclareMathOperator*{\argmin}{arg\,min}

\usepackage{algorithmic}
\usepackage{graphicx}
\DeclareGraphicsExtensions{.pdf,.jpeg,.png}

\usepackage{textcomp}
\usepackage{xcolor}
\usepackage{multicol}
\usepackage{multirow}
\usepackage{afterpage}
\usepackage{siunitx}
\usepackage{nccmath}
\usepackage{array}
\usepackage{makecell}
\usepackage{hhline}
\usepackage{booktabs}
    
\newcommand{\thickhline}{
    \noalign {\ifnum 0=`}\fi \hrule height 1pt
    \futurelet \reserved@a \@xhline
}

\makeatletter
\newcommand{\vast}{\bBigg@{4}}
\newcommand{\Vast}{\bBigg@{5}}
\makeatother

\begin{document}

\title{Optimal Parameter Inflation to Enhance the Availability of Single-Frequency GBAS for Intelligent Air Transportation}

\author{Halim~Lee, Sam~Pullen, Jiyun~Lee,~\IEEEmembership{Member,~IEEE,}
Byungwoon~Park,~\IEEEmembership{Member,~IEEE,}
Moonseok~Yoon,
and~Jiwon~Seo,~\IEEEmembership{Member,~IEEE}
\thanks{Manuscript received January 26, 2021; revised January 00, 2022; }
\thanks{This work was supported in part by the Unmanned Vehicles Core Technology Research and Development Program through the National Research Foundation of Korea (NRF) and the Unmanned Vehicle Advanced Research Center (UVARC) funded by the Ministry of Science and ICT, Republic of Korea
(2020M3C1C1A01086407), and in part by the Institute of Information and Communications Technology Planning and Evaluation (IITP) grant funded by the Korea government (KNPA) (2019-0-01291). \textit{(Corresponding Author: Jiwon Seo.)}}
\thanks{H. Lee and J. Seo are with the School of Integrated Technology, Yonsei University, Incheon 21983, Republic of Korea (e-mail: halim.lee@yonsei.ac.kr; jiwon.seo@yonsei.ac.kr).}
\thanks{S. Pullen is with the Department of Aeronautics and Astronautics, Stanford University, Stanford, CA 94305, USA (e-mail: spullen@stanford.edu).}
\thanks{J. Lee is with the Department of Aerospace Engineering, Korea Advanced Institute of Science and Technology, Daejeon 34141, Republic of Korea (e-mail: jiyunlee@kaist.ac.kr).}
\thanks{B. Park is with the Department of Aerospace Engineering and the Department of Convergence Engineering for Intelligent Drone, Sejong University, Seoul 05006, Republic of Korea (e-mail: byungwoon@sejong.ac.kr).}
\thanks{M. Yoon is with the Korea Aerospace Research Institute, Daejeon 34133, Republic of Korea (e-mail: msyoon@kari.re.kr).}
}

%
%

\markboth{T-ITS-21-01-0209.R2}%
{Lee \MakeLowercase{\textit{et al.}}: Optimal Parameter Inflation to Enhance the Availability of Single-Frequency GBAS for Intelligent Air Transportation}
%



\maketitle

\begin{abstract} 
Ground-based Augmentation System (GBAS) augments Global Navigation Satellite Systems (GNSS) to support the precision approach and landing of aircraft. To guarantee integrity, existing single-frequency GBAS utilizes position-domain geometry screening to eliminate potentially unsafe satellite geometries by inflating one or more broadcast GBAS parameters. However, GBAS availability can be drastically impacted in low-latitude regions where severe ionospheric conditions have been observed. Thus, we developed a novel geometry-screening algorithm in this study to improve GBAS availability in low-latitude regions. Simulations demonstrate that the proposed method can provide 5--8 percentage point availability enhancement of GBAS at Galeão airport near Rio de Janeiro, Brazil, compared to existing methods.  
\end{abstract}


%
\IEEEpeerreviewmaketitle

\section{Introduction}
\IEEEPARstart{G}{LOBAL} Navigation Satellite Systems (GNSS), such as the U.S. Global Positioning System (GPS), are essential components in aviation infrastructure. Future intelligent air transportation systems will rely heavily on GNSS for navigation and surveillance \cite{Walter08:Worldwide, Prandini11:Toward, Seo14:Future, Sandamali20:Flight, Scala20:Tackling}; however, GNSS alone cannot meet the stringent integrity requirements for civil aviation applications. To compensate this, Satellite-based Augmentation Systems (SBAS) \cite{Enge96:Wide, Neish19:Design, Yoon20:An} and Ground-based Augmentation Systems (GBAS) \cite{Enge99:Local, Gerbeth19:Satellite} have been developed and deployed. SBAS and GBAS monitor GNSS signals, generate differential corrections and integrity information, and broadcast them to users. Users with certified GNSS aviation receivers can utilize GNSS signals along with the received corrections and integrity information to perform aviation applications that would be difficult with GNSS alone. While SBAS covers large regions, GBAS covers local areas around airports and supports flight operations closer to the ground. Here, we focus on existing single-frequency GBAS that guides aircraft in poor visibility down to the minimum 200-ft decision height of a Category I precision approach.

The ionosphere is the greatest source of error for GNSS, and
integrity threats arising from anomalous ionospheric behavior \cite{Mendillo06:Storms, Saito17:Evaluation, Seo11:Availability, Sun21:Markov} to single-frequency GBAS users must be mitigated \cite{Lee17:Monitoring, Lee17:Optimized, Shively08:Safety, Mayer09:Ionosphere}. One challenging phenomenon is the sharp electron density gradients within the ionosphere \cite{Rungraengwajiake15:Ionospheric, Saito12:Precise, Saito17:Ionospheric, Sanchez-Naranjo17:A, Fujita10:Determination} that may occur under active ionospheric gradient conditions \cite{Datta-Barua10:Ionospheric, Lee07:Assessment, Yoon19:Assessment}. A single-frequency GBAS ground facility cannot directly measure ionospheric spatial gradients; thus, these systems must assume that the worst-case ionospheric gradient historically observed in a given region is always present when using geometry screening methods \cite{Lee11:Ionospheric, Seo12:Targeted} in order to satisfy the stringent aviation integrity requirements. 
\begin{table*} 
\centering
\caption{Comparison of existing and proposed position-domain geometry screening algorithms}
\label{tab:CompMethods}

\vspace{-4mm}
\begin{center}
{\renewcommand{\arraystretch}{1.4}
 \begin{tabular}[c]{>{\centering\arraybackslash}m{3cm}>{\arraybackslash}m{6.5cm}>{\arraybackslash}m{6.5cm}}
 \toprule
    \thead{Algorithm} & \thead{Approach} & \thead{Characteristics}\\
\midrule
{$\sigma_\mathrm{vig}$ inflation \cite{Lee11:Ionospheric}} 
& {Repeatedly inflates $\sigma_\mathrm{vig}$ in small amounts and checks whether the VPLs of all potentially unsafe satellite geometries exceed the VAL.} 
& {Simple to implement and fast to obtain an inflated $\sigma_\mathrm{vig}$ but with reduced GBAS availability} \\
{Targeted inflation \cite{Seo12:Targeted}} 
& {Inflates two satellite-specific parameters, $\sigma_{\mathrm{pr}_\mathrm{gnd}}$ and $P_k$, using linear programs.} 
& {Significant GBAS availability benefit over $\sigma_\mathrm{vig}$ inflation at mid-latitude}\\
{Optimal $\sigma_{\mathrm{pr}_\mathrm{gnd}}$ inflation (proposed)} 
& {Inflates the satellite-specific parameter, $\sigma_{\mathrm{pr}_\mathrm{gnd}}$, using a linear program.} 
& {Better GBAS availability at low-latitude and faster computation time than targeted inflation}\\
 \bottomrule
\end{tabular}}
\end{center}
\end{table*}

In this work, we studied the limitations of existing geometry screening methods \cite{Lee11:Ionospheric, Seo12:Targeted} for low-latitude regions. We found, for the first time, the performance degradation of the targeted inflation algorithm \cite{Seo12:Targeted} in low-latitude regions. To overcome the limitations of the existing methods, we propose a novel geometry-screening algorithm, namely optimal $\sigma_{\mathrm{pr}_\mathrm{gnd}}$ inflation (Table~\ref{tab:CompMethods}). The proposed algorithm demonstrates better performance for the majority of nighttime epochs than previous geometry screening algorithms. 

\section{Background}
\label{sec:Background}

GBAS must ensure aviation integrity even when the weakest satellite signal pair is impacted by the worst-case ionospheric gradient recorded (i.e., a ``worst-case impact''). A given distribution of satellites in view (i.e., ``satellite geometry'') on a certain epoch may not be safe to use under a worst-case impact. If such a satellite geometry exists, GBAS must prevent the use of potentially hazardous satellite geometry during aviation. This can be achieved by the \textit{position-domain geometry screening} algorithm currently utilized by operational single-frequency GBAS supporting Category I precision approaches.

GBAS avionics are already installed on certain aircraft, and communication message formats between the user and GBAS ground facility are standardized \cite{RTCA17:GNSS}. Without a dedicated message, it is difficult to notify users of potentially unsafe satellite geometries. 
There are at least two well-known position-domain geometry-screening algorithms described in the literature \cite{Lee11:Ionospheric, Seo12:Targeted} to resolve this problem, which are summarized in Table~\ref{tab:CompMethods}. 

The idea of position-domain geometry screening is to inflate the \textit{vertical protection levels} (VPLs), which are the bounds of vertical position errors, of potentially unsafe subset geometries above the \textit{vertical alert limit} (VAL) so that such geometries are excluded from being used by the aircraft. The VPL of each subset geometry is calculated as follows \cite{RTCA17:Minimum}:

\begin{fleqn}[\parindent]
\begin{equation} \label{eq:eq1}
    \mathrm{VPL}_\mathrm{H0} = K_\mathrm{ffmd} \sqrt{\sum_{j=1}^{N_\mathrm{U}} S_{\mathrm{U},\mathrm{vert},j}^2 \sigma_j^2}
\end{equation}
\end{fleqn}

\begin{fleqn}[\parindent] 
\begin{equation} \label{eq:eq2}
\begin{split}
    \mathrm{VPL}_\mathrm{eph} = \max_{k} \Bigg\{ & \big| S_{\mathrm{U},\mathrm{vert},k} \big| x_\mathrm{aircraft} P_k + \\
    & K_{\mathrm{md}_\mathrm{eph}} \sqrt{\sum_{j=1}^{N_\mathrm{U}} S_{\mathrm{U},\mathrm{vert},j}^2 \sigma_j^2} \Bigg\}
\end{split}
\end{equation}
\end{fleqn}

\begin{fleqn}[\parindent] 
\begin{equation} \label{eq:eq3}
    \mathrm{VPL} = \max\left\{ \mathrm{VPL}_\mathrm{H0}, \; \mathrm{VPL}_\mathrm{eph} \right\}
\end{equation}
\end{fleqn}
where $N_\mathrm{U}$ is the number of satellites in a given subset geometry; $K_\mathrm{ffmd}$ and $K_{\mathrm{md}_\mathrm{eph}}$ are constants derived from the probability of fault-free missed detection and the probability of missed detection with an ephemeris error, respectively; $x_\mathrm{aircraft}$ is the separation (i.e., absolute distance) between an airplane and GBAS ground facility; $P_k$ is the ephemeris error decorrelation parameter for satellite $k$; and $S_{\mathrm{U},\mathrm{vert},j}$ is the vertical position component of the weighted least-squares projection matrix of the given subset geometry, which is a function of the variance $\sigma_j^2$ of a normal distribution that overbounds the true post-correction range error of satellite $j$. The variance $\sigma_j^2$ is a function of $\sigma_{\mathrm{pr}_\mathrm{gnd},j}^2$ and $\sigma_\mathrm{vig}$, as in (\ref{eq:eq4}) and (\ref{eq:eq5}), which are calculated and broadcast by the GBAS ground facility.

\begin{fleqn}[\parindent] 
\begin{equation} 
\label{eq:eq4}
    \sigma_j^2 = \sigma_{\mathrm{pr}_\mathrm{gnd},j}^2 + \sigma_{\mathrm{tropo},j}^2 + \sigma_{\mathrm{pr}_\mathrm{air},j}^2 + \sigma_{\mathrm{iono},j}^2
\end{equation}
\end{fleqn}
    
\begin{fleqn}[\parindent] 
\begin{equation} 
\label{eq:eq5}
    \sigma_{\mathrm{iono},j}^2 = F_j \sigma_\mathrm{vig} (x_\mathrm{aircraft} + 2 \tau v_\mathrm{aircraft})
\end{equation}
\end{fleqn}
where $\sigma_{\mathrm{pr}_\mathrm{gnd},j}^2$ is the variance of the ground error term for satellite $j$; $\sigma_{\mathrm{tropo},j}^2$ is the variance of the tropospheric error term for satellite $j$; $\sigma_{\mathrm{pr}_\mathrm{air},j}^2$ is the variance of the fault-free airborne receiver measurement error term for satellite $j$; $\sigma_{\mathrm{iono},j}^2$ is the variance of the ionospheric error term for satellite $j$;  $F_j$ is the ionospheric thin shell model obliquity factor for satellite $j$; $\sigma_\mathrm{vig}$ is the standard deviation of residual ionospheric uncertainty (``vig'' stands for vertical ionospheric gradient); $\tau$ is the time constant of the single-frequency carrier-smoothing filter; and $v_\mathrm{aircraft}$ is the horizontal approach velocity of the airplane. (See \cite{Lee11:Ionospheric, Murphy10:GBAS, Park10:Enabling} for the detailed error models.)

\begin{figure}
    \centering
    \includegraphics[width=0.95\linewidth]{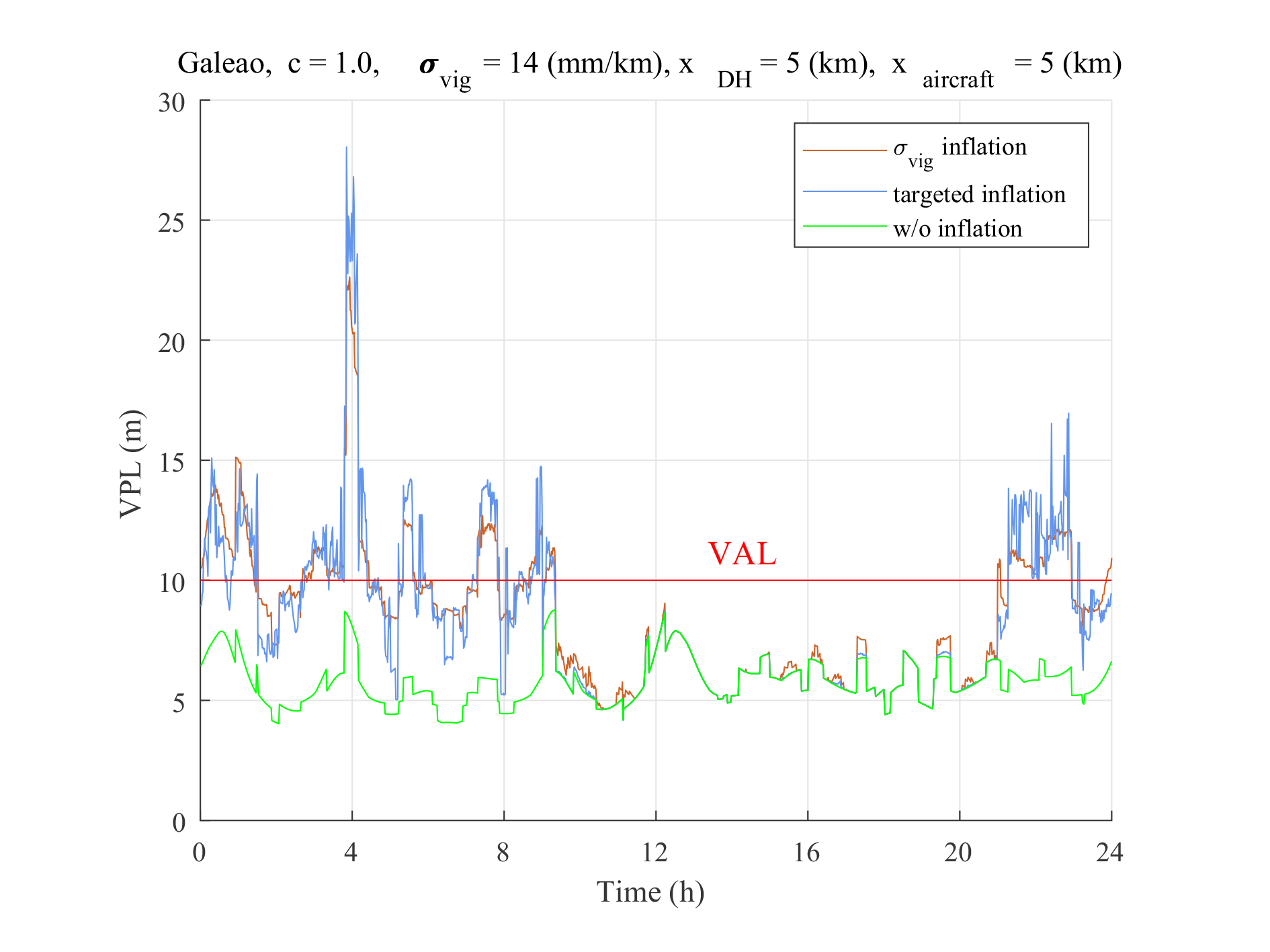}
    \caption{Increased VPL for the all-in-view satellite geometry due to position-domain geometry-screening algorithms (i.e., $\sigma_\mathrm{vig}$ inflation and targeted inflation) at Galeão airport in Brazil. GBAS is unavailable when VPL exceeds VAL. The 10-m VAL is from Fig.~\ref{fig:VAL_TEL} when $x_\mathrm{DH}$ and $x_\mathrm{aircraft}$ are the same. Local nighttime with more severe ionospheric threats corresponds to 0 to 9 h and 21 to 24 h in UT (shown on the x-axis).}
    \label{fig:ExistingMethods}
\end{figure}

However, when the existing geometry screening algorithms \cite{Lee11:Ionospheric, Seo12:Targeted} are applied to low-latitude regions, where the observed worst-case ionospheric gradient is much greater than that of  mid-latitude \cite{Pradipta16:Assessing}, the availability of GBAS is greatly impacted. 
In Fig.~\ref{fig:ExistingMethods}, the nominal VPL of all-in-view satellite geometry at Galeão airport in Brazil without geometry screening (green curve) is always less than VAL. However, the integrity threat due to the sharp ionospheric gradient is not mitigated in this case; thus, navigation integrity cannot be guaranteed. When a geometry-screening algorithm such as $\sigma_\mathrm{vig}$ inflation \cite{Lee11:Ionospheric} or targeted inflation \cite{Seo12:Targeted} is applied with the low-latitude ionospheric threat model, integrity is assured, but the resultant VPLs increase due to parameter inflation. Notably, VPLs in both inflation methods (brown and blue curves) exceed VAL at many local nighttime epochs (i.e., 0 to 9 h and 21 to 24 h in Fig.~\ref{fig:ExistingMethods}), for which GBAS is unavailable.

\section{Optimal $\sigma_{\mathrm{pr}_\mathrm{gnd}}$ Inflation Algorithm} 
\label{sec:ProposedMethod}

\subsection{Optimization Problem Formulation} 
\label{sec:OptProbForm}

The position-domain geometry-screening problem can be formulated as an optimization problem, as shown in (\ref{eqn:OriginalProblem}) \cite{Seo12:Targeted}. Each unsafe subset geometry is screened by the constraint, while the objective function minimizes the VPL of the all-in-view geometry.

\begin{fleqn}[\parindent] 
\begin{equation} 
\label{eqn:OriginalProblem}
\begin{split}
    \text{Minimize} \quad & \max \left\{ \mathrm{VPL}_\mathrm{H0},  \mathrm{VPL}_\mathrm{eph} \right\} \\
    & \text{for the all-in-view geometry} \\
    \text{Subject to} \quad & \max \left\{ \mathrm{VPL}_\mathrm{H0}, \mathrm{VPL}_\mathrm{eph} \right\}  \geq \mathrm{VAL} \\
    & \text{for each unsafe subset geometry}
\end{split}
\end{equation}
\end{fleqn}
Because GBAS operates in real-time and its computing resources are limited, a technique that directly solves the nonlinear optimization problem in (\ref{eqn:OriginalProblem}) is undesirable. Thus, the targeted inflation \cite{Seo12:Targeted} and our optimal $\sigma_{\mathrm{pr}_\mathrm{gnd}}$ inflation algorithms reduce the nonlinear optimization problem in (\ref{eqn:OriginalProblem}) to linear programs (LPs) to find a good suboptimal solution. 
Further, our optimal $\sigma_{\mathrm{pr}_\mathrm{gnd}}$ inflation provides a better suboptimal solution than that of targeted inflation for low-latitude regions.

The optimal $\sigma_{\mathrm{pr}_\mathrm{gnd}}$ inflation algorithm focuses on the satellite-specific parameter, $\sigma_{\mathrm{pr}_\mathrm{gnd},j}$, for each satellite $j$ among the broadcast parameters that can be inflated by the GBAS ground facility. 
Although the targeted inflation algorithm \cite{Seo12:Targeted} inflates both $\sigma_{\mathrm{pr}_\mathrm{gnd},j}$ and $P_k$,
we found that the LP for the inflation of $P_k$ was often infeasible at low-latitude under nighttime conditions.

The $P_k$ in the $\mathrm{VPL}_\mathrm{eph}$ equation is not inflated by the optimal $\sigma_{\mathrm{pr}_\mathrm{gnd}}$ inflation; thus, the objective of the original optimization problem in (\ref{eqn:OriginalProblem}) is changed to the objective in (\ref{eqn:VPL_H0_Opt}).

\begin{fleqn}[\parindent] 
\begin{equation} 
\label{eqn:VPL_H0_Opt}
\begin{split}
    \text{Minimize} \quad & \mathrm{VPL}_\mathrm{H0} \\
    & \text{for the all-in-view geometry} \\
    \text{Subject to} \quad & \mathrm{VPL}_\mathrm{H0} \geq \mathrm{VAL} \; \; \text{or} \; \; \mathrm{VPL}_\mathrm{eph} \geq \mathrm{VAL} \\
    & \text{for each unsafe subset geometry} \\
\end{split}
\end{equation}
\end{fleqn}
Using (\ref{eq:eq1}) and (\ref{eq:eq2}), the new optimization problem in (\ref{eqn:VPL_H0_Opt}) is expressed as (\ref{eqn:OptInfInit}).

\begin{fleqn}[\parindent] 
\begin{equation} 
\label{eqn:OptInfInit}
\begin{split}
    \underset{\sigma_i^2}{\text{Minimize}}  \quad & K_\mathrm{ffmd} \sqrt{\sum_{i=1}^N S_{\mathrm{vert},i}^2 \sigma_i^2} \\
    & \text{for the all-in-view geometry} \\
    \text{Subject to}  \quad & K_\mathrm{ffmd} \sqrt{\sum_{j=1}^{N_\mathrm{U}} S_{\mathrm{U},\mathrm{vert},j}^2 \sigma_j^2} \geq \mathrm{VAL} \; \; \text{or}\\
     \quad & \max_{k} \left( \big| S_{\mathrm{U}, \mathrm{vert}, k} \big| P_k \right) x_\mathrm{aircraft} \, + \\
    & \quad \quad K_{\mathrm{md}_\mathrm{eph}} \sqrt{\sum_{j=1}^{N_\mathrm{U}} S_{\mathrm{U},\mathrm{vert},j}^2 \sigma_j^2} \geq \mathrm{VAL} \\
    & \text{for each unsafe subset geometry} \\
\end{split}
\end{equation}
\end{fleqn}
where $N$ is the number of satellites in the all-in-view geometry and $S_{\mathrm{vert},i}$ is the vertical position component of the weighted least-squares projection matrix of the all-in-view geometry.
Since the values of $x_\mathrm{aircraft}$ and $K_{\mathrm{md}_\mathrm{eph}} \sqrt{\sum_{j=1}^{N_\mathrm{U}} S_{\mathrm{U},\mathrm{vert},j}^2 \sigma_j^2}$ in (\ref{eq:eq2}) are independent of $k$, they are outside of the $\max_{k} \left( \cdot \right)$ operator in (\ref{eqn:OptInfInit}).

Further, the two constraint equations in (\ref{eqn:OptInfInit}) are combined into a single constraint equation in (\ref{eqn:OptInfConstraint}) to formulate the optimization problem in the standard LP form, without having an ``or'' constraint expression.

\begin{fleqn}[\parindent] 
\begin{equation} 
\label{eqn:OptInfConstraint}
\begin{split}
    - & \sum_{j=1}^{N_\mathrm{U}} S_{\mathrm{U},\mathrm{vert},j}^2 \sigma_j^2 \leq \\
    & \max \Bigg\{ -\left( \frac{\mathrm{VAL}}{K_\mathrm{ffmd}} \right)^2, \\
    & \quad \quad \; \; \, -\left(\frac{\mathrm{VAL} - \max_{k} \left( \big| S_{\mathrm{U}, \mathrm{vert}, k} \big| P_k \right) x_\mathrm{aircraft}}{K_{\mathrm{md}_\mathrm{eph}}}\right)^2 \Bigg\}
\end{split}
\end{equation}
\end{fleqn}
$P_k$ is constant and is not inflated; thus, the right-hand side of (\ref{eqn:OptInfConstraint}) can be calculated as a constant value if the value of $S_{\mathrm{U},\mathrm{vert},k}$, which is a function of $\sigma_j^2$, is calculated using the nominal $\sigma_{\mathrm{pr}_\mathrm{gnd},j}$ value without inflation. 
This approach of fixing $S_{\mathrm{U},\mathrm{vert},k}$ as a constant is suboptimal but necessary to reduce (\ref{eqn:OptInfConstraint}) to an LP. Targeted inflation also uses the same approach. (Detailed discussion of the effect of this approach is given in Section~\ref{sec:Adjustment}). 
Then, the left-hand side of (\ref{eqn:OptInfConstraint}) is a linear combination of $\sigma_j^2$, as $S_{\mathrm{U},\mathrm{vert},j}^2$ is constant. The objective function in (\ref{eqn:OptInfInit}) is equivalent to minimizing $\sum_{i=1}^{N} S_{\mathrm{vert},i}^2 \sigma_i^2$, which is also a linear combination of $\sigma_j^2$ since $S_{\mathrm{vert},i}^2$ is constant.

The GBAS interface control document (ICD) \cite{RTCA17:GNSS} specifies the maximum possible $\sigma_{\mathrm{pr}_\mathrm{gnd}}$ value, i.e., 5.08 m. Thus, another linear constraint equation in (\ref{eq:eq11}) for each satellite $i$ also needs to be considered.

\begin{fleqn}[\parindent] 
\begin{equation} 
\label{eq:eq11}
\begin{split}
    &\sigma_{\mathrm{pr}_\mathrm{gnd},i,0}^2 + \sigma_{\mathrm{tropo},i}^2 + \sigma_{\mathrm{pr}_\mathrm{air},i}^2 + \sigma_{\mathrm{iono},i}^2 \leq \\
    & \quad \quad \quad \quad \sigma_i^2 \leq 5.08^2 + \sigma_{\mathrm{tropo},i}^2 + \sigma_{\mathrm{pr}_\mathrm{air},i}^2 + \sigma_{\mathrm{iono},i}^2
\end{split}
\end{equation}
\end{fleqn}
where $\sigma_{\mathrm{pr}_\mathrm{gnd},i,0}$ is the nominal $\sigma_{\mathrm{pr}_\mathrm{gnd},i}$ value without inflation. 

Finally, the LP for optimal $\sigma_{\mathrm{pr}_\mathrm{gnd}}$ inflation is formulated as in (\ref{eqn:FinalLP}) by combining (\ref{eqn:OptInfInit})--(\ref{eq:eq11}).
The objective functions of (\ref{eqn:OptInfInit}) and (\ref{eqn:FinalLP}) are different, but $\argmin_{\sigma_i^2} K_\mathrm{ffmd} \sqrt{\sum_{i=1}^N S_{\mathrm{vert},i}^2 \sigma_i^2}$ is equal to $\argmin_{\sigma_i^2} \sum_{i=1}^N S_{\mathrm{vert},i}^2 \sigma_i^2$ because $K_\mathrm{ffmd}$ is a constant.

\begin{fleqn}[\parindent] 
\begin{equation} 
\label{eqn:FinalLP}
\begin{split}
    \underset{\sigma_i^2}{\text{Minimize}}  \quad & \sum_{i=1}^N S_{\mathrm{vert},i}^2 \sigma_i^2 \\
    & \text{for the all-in-view geometry} \\
    \text{Subject to}  \quad & -\sum_{j=1}^{N_\mathrm{U}} S_{\mathrm{U},\mathrm{vert},j}^2 \sigma_j^2 \leq \\
    \max \Bigg\{ - & \left( \frac{\mathrm{VAL}}{K_\mathrm{ffmd}} \right)^2, \\
     - & \left(\frac{\mathrm{VAL}-\max_{k} \left( \big| S_{\mathrm{U}, \mathrm{vert}, k} \big| P_k \right) x_\mathrm{aircraft}}{K_{\mathrm{md}_\mathrm{eph}}}\right)^2 \Bigg\} \\
    & \text{for each unsafe subset geometry} \\
    &\sigma_{\mathrm{pr}_\mathrm{gnd},i,0}^2 + \sigma_{\mathrm{tropo},i}^2 + \sigma_{\mathrm{pr}_\mathrm{air},i}^2 + \sigma_{\mathrm{iono},i}^2 \leq \\
    & \quad \; \sigma_i^2 \leq 5.08^2 + \sigma_{\mathrm{tropo},i}^2 + \sigma_{\mathrm{pr}_\mathrm{air},i}^2 + \sigma_{\mathrm{iono},i}^2 \\
    & \text{for each satellite} \; i
\end{split}
\end{equation}
\end{fleqn}
Once the optimal $\sigma_i^2$ value for each satellite is obtained by solving this LP, the optimal $\sigma_{\mathrm{pr}_\mathrm{gnd}}$ value for satellite $i$ is calculated by (\ref{eq:eq13}).

\begin{fleqn}[\parindent] 
\begin{equation} 
\label{eq:eq13}
\begin{split}
    \sigma_{\mathrm{pr}_\mathrm{gnd}, i} = \sqrt{\sigma_i^2 - \sigma_{\mathrm{tropo},i}^2 - \sigma_{\mathrm{pr}_\mathrm{air},i}^2 - \sigma_{\mathrm{iono},i}^2}
\end{split}
\end{equation}
\end{fleqn}

The expression of ``each unsafe subset geometry'' in (\ref{eqn:FinalLP}) implicitly contains $x_\mathrm{DH}$ and $x_\mathrm{aircraft}$. 
$x_\mathrm{DH}$ is the horizontal distance between the GBAS ground facility and the airplane when it reaches the 200-ft minimum decision height for a Category-I precision approach. The $x_\mathrm{DH}$ locations are specified for each airport. 
Although $x_\mathrm{DH}$ is not explicitly shown in (\ref{eqn:FinalLP}), the tolerable error limit (TEL) is a function of $x_\mathrm{DH}$ and $x_\mathrm{aircraft}$, as shown in Fig.~\ref{fig:VAL_TEL}, and the maximum ionosphere-induced error in vertical (MIEV) must be compared to TEL in order to identify unsafe subset geometries. 
(See \cite{Lee11:Ionospheric} for the TEL and MIEV calculations. GBAS ground facility calculates TEL, MIEV, and VPL for geometry screening, whereas aircraft calculates VPL only.)

\begin{figure}
    \centering
    \includegraphics[width=0.85\linewidth]{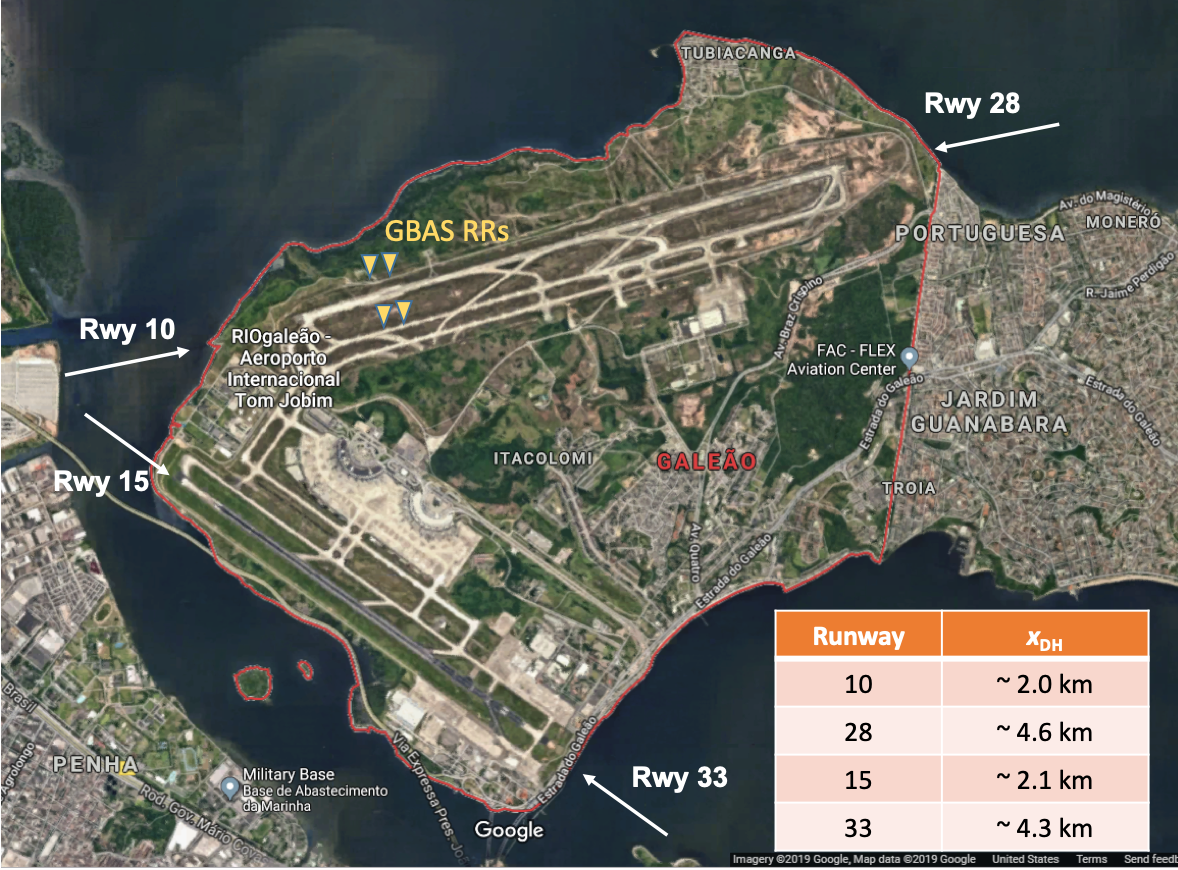}
    \caption{$x_\mathrm{DH}$ values for Galeão airport. Each runway has a known $x_\mathrm{DH}$ location where an airplane reaches the minimum 200 ft decision height for a Category I precision approach.}
    \label{fig:GaleaoAirport}
\end{figure}

\begin{figure}
    \centering
    \includegraphics[width=0.85\linewidth]{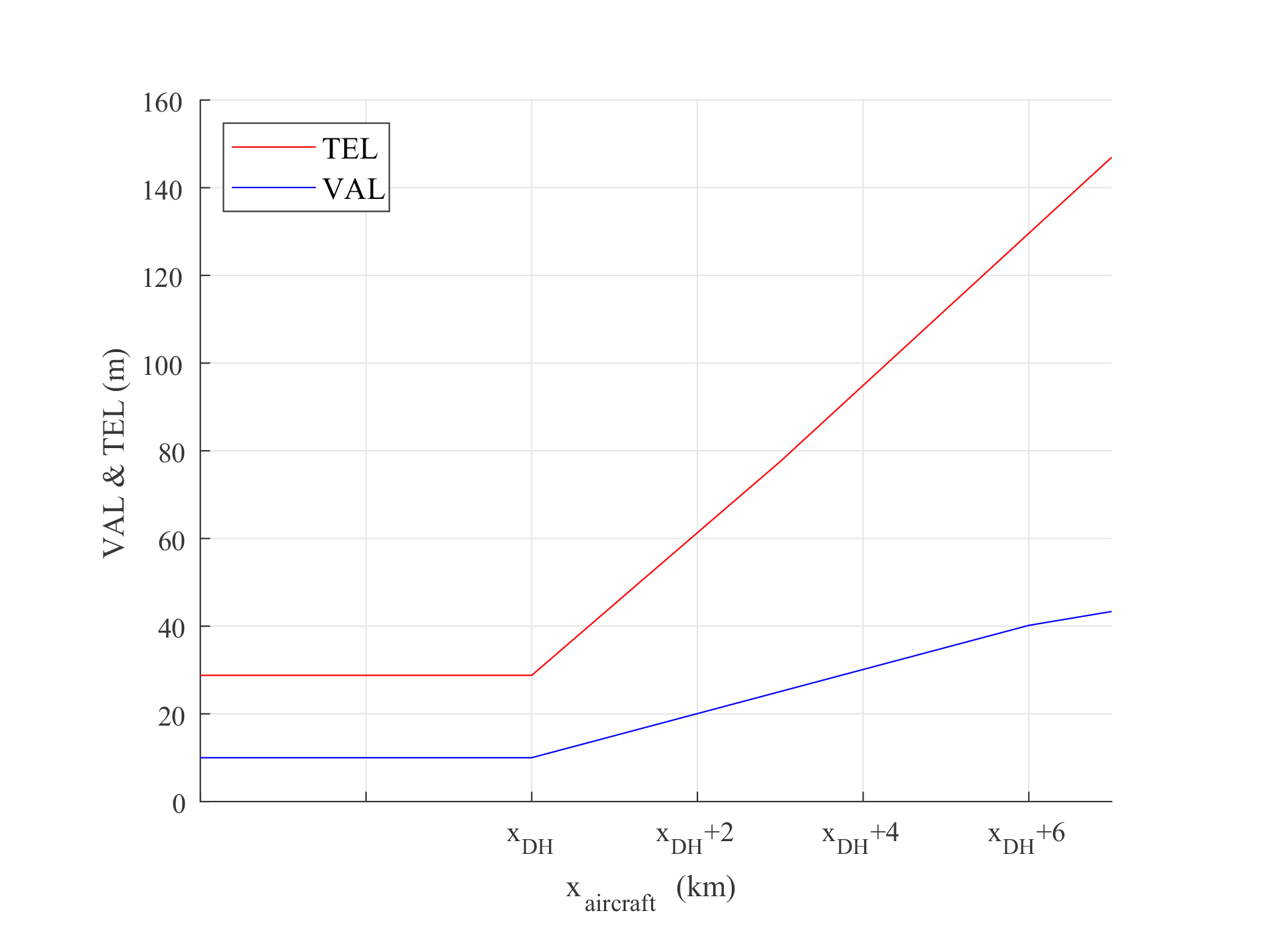}
    \caption{VAL and TEL according to $x_\mathrm{aircraft}$ and $x_\mathrm{DH}$. These values were also used in \cite{Lee11:Ionospheric} and \cite{Seo12:Targeted}. For example, if an airplane at a 5 km distance from the GBAS ground facility (i.e., $x_\mathrm{aircraft}$ is 5 km)  approaches to the runway 10 in Fig.~\ref{fig:GaleaoAirport}, its VAL and TEL are approximately 25 m and 78 m, respectively, because $x_\mathrm{DH}$ of the runway is 2 km and $x_\mathrm{aircraft}$ is $x_\mathrm{DH} + 3$ km.}
    \label{fig:VAL_TEL}
\end{figure}

To develop GBAS software to ensure the integrity at Galeão airport for any airplane approaching any runway end, the four $x_\mathrm{DH}$ values shown in Fig.~\ref{fig:GaleaoAirport} must be considered when the constraints in (\ref{eqn:FinalLP}) are generated. 
If the same GBAS software is to be deployed at any airport, all possible $x_\mathrm{DH}$ values should be considered for geometry screening. Here, we considered $x_\mathrm{DH}$ values from 0 to 6 km (1 km increments) and $x_\mathrm{aircraft}$ from $x_\mathrm{DH}$ to $x_\mathrm{DH}$ + 7 km (1 km increments) when generating the constraints in (\ref{eqn:FinalLP}). These considerations should be able to cover all possible $x_\mathrm{DH}$ and $x_\mathrm{aircraft}$ values for almost any airport with a margin.

An important advantage of using optimal $\sigma_{\mathrm{pr}_\mathrm{gnd}}$ inflation over targeted inflation is that the LP in (\ref{eqn:FinalLP}) needs to be solved only once for each epoch, even when multiple $x_\mathrm{aircraft}$ and $x_\mathrm{DH}$ values are considered. Targeted inflation \cite{Seo12:Targeted} requires solving a maximum of two LPs for each combination of $x_\mathrm{aircraft}$ and $x_\mathrm{DH}$ for which inflation is required. Hence, if eight $x_\mathrm{aircraft}$ and seven $x_\mathrm{DH}$ values are considered, as in this study, the theoretical maximum number of LPs to be solved for one epoch is 112. In practice, certain combinations of $x_\mathrm{aircraft}$ and $x_\mathrm{DH}$ values do not require parameter inflation. 
For example, the targeted inflation solves 97 LPs for the epoch, corresponding to 2 h 52 min in Fig.~\ref{fig:ExistingMethods}, while the optimal $\sigma_{\mathrm{pr}_\mathrm{gnd}}$ inflation solves only one LP although the size of this LP is larger than the size of each LP of targeted inflation. The computational times for targeted inflation and optimal $\sigma_{\mathrm{pr}_\mathrm{gnd}}$ inflation at this epoch were 3.58 s and 1.65 s, respectively, using a Matlab script running on a PC with an Intel Xeon E3-1270 CPU. Inflation factors are updated every minute; thus, both methods are potentially suitable for real-time GBAS operations. As both methods require additional operations besides solving LPs, their computational times do not differ by a factor of hundred. Nevertheless, optimal $\sigma_{\mathrm{pr}_\mathrm{gnd}}$ inflation requires perceptibly less computational time than targeted inflation. Considering that the simplest algorithm without solving any optimization problem (i.e., $\sigma_\mathrm{vig}$ inflation) took 0.68 s for the same epoch, the proposed algorithm solving one LP is reasonably fast.

\subsection{$\sigma_{\mathrm{pr}_\mathrm{gnd}}$ Inflation and Adjustment}
\label{sec:Adjustment}

The LP in (\ref{eqn:FinalLP}) is obtained after making $S_{\mathrm{U},\mathrm{vert},j}$ constant using the nominal $\sigma_{\mathrm{pr}_\mathrm{gnd},j}$ value. When $\sigma_{\mathrm{pr}_\mathrm{gnd},j}$ is inflated by solving (\ref{eqn:FinalLP}), the actual $S_{\mathrm{U},\mathrm{vert},j}$ value changes accordingly. Hence, the constraints in (\ref{eqn:FinalLP}) may not be satisfied with the updated $S_{\mathrm{U},\mathrm{vert},j}$ value. There are at least two ways to resolve this problem.

First, we may consider an iterative approach. After obtaining the inflated $\sigma_{\mathrm{pr}_\mathrm{gnd},j}$ by solving (\ref{eqn:FinalLP}), $S_{\mathrm{U},\mathrm{vert},j}$ can be updated using the inflated $\sigma_{\mathrm{pr}_\mathrm{gnd},j}$. Then, the optimization problem in (\ref{eqn:FinalLP}) with the updated $S_{\mathrm{U},\mathrm{vert},j}$ is solved again to update the inflated $\sigma_{\mathrm{pr}_\mathrm{gnd},j}$. This process can be repeated until the constraints in (\ref{eqn:FinalLP}) are satisfied. However, the iterations  require more computational time, and we could not prove that it converges mathematically under all conditions.

The second option, which we used in this study, is to increase the inflated $\sigma_{\mathrm{pr}_\mathrm{gnd},j}$ that is obtained by solving the LP in (\ref{eqn:FinalLP}) until the constraints are satisfied. As with the first option, the constraint check process remains the same. However, when the constraints are not met, where the first option solves the LP again, the second option increases $\sigma_{\mathrm{pr}_\mathrm{gnd},j}$ by a small amount (0.02 m in this study, as it is the minimum resolution of the broadcast $\sigma_{\mathrm{pr}_\mathrm{gnd},j}$ \cite{RTCA17:GNSS}). The second option requires less computational time, and convergence is not a limitation because the constraints will eventually be satisfied if $\sigma_{\mathrm{pr}_\mathrm{gnd},j}$ continues to increase. Provided that the constraints in (\ref{eqn:FinalLP}), which are the same as the constraints of the original nonlinear optimization problem in (\ref{eqn:OriginalProblem}), are satisfied, the navigation integrity is guaranteed regardless of the suboptimality of the reduced LP in (\ref{eqn:FinalLP}).

There are two possible stop conditions for the second option. The iterative increment of $\sigma_{\mathrm{pr}_\mathrm{gnd},j}$ stops when the constraints in (\ref{eqn:FinalLP}) are satisfied.
It is theoretically possible that the constraints would not be satisfied even with the maximum possible $\sigma_{\mathrm{pr}_\mathrm{gnd},j}$ value (i.e., 5.08 m) for all satellites and the algorithm would stop without satisfying the constraints. However, we did not observe this issue in practice.  The largest inflation of $\sigma_{\mathrm{pr}_\mathrm{gnd},j}$ values occurs when all subset geometries are unsafe for use. This can occur if the worst-case ionospheric gradient in the threat model is very large. To address this, we assessed a hypothetical situation with a worst-case gradient of 10,000 mm/km, which is more than 10 times larger than the observed worst-case gradient of 850.7 mm/km \cite{Yoon19:Assessment}. In this case, as expected, all subset geometries of all time epochs were unsafe for use, which is the worst possible situation for geometry screening. However, each subset geometry was successfully screened by $\sigma_{\mathrm{pr}_\mathrm{gnd},j}$ values less than their maximum possible value of 5.08 m.

Unlike in \cite{Seo12:Targeted}, wherein $\sigma_{\mathrm{pr}_\mathrm{gnd},j}$ is adjusted by the same amount for all satellites in the same subset geometry, we instead adjusted $\sigma_{\mathrm{pr}_\mathrm{gnd},j}$ for each satellite in turn (in no particular order) to minimize unnecessary inflation. When we compared the performance between targeted inflation and optimal $\sigma_{\mathrm{pr}_\mathrm{gnd}}$ inflation, we used the same adjustment algorithm (i.e., the second option) for both methods to enable a fair comparison.

\section{Performance Analysis} 
\label{sec:Performance}

\subsection{Experimental Design}

The performances of each geometry-screening algorithm in Table~\ref{tab:CompMethods} were compared for both low- and mid-latitude regions. 
A number of airports---Galeão airport in Brazil, New Ishigaki airport in Japan, Chennai airport in India, and Memphis airport in the U.S.---were considered. Galeão, Ishigaki, and Chennai airports are in low-latitude regions, while Memphis airport is in a mid-latitude region.

Yoon \textit{et al.} \cite{Yoon19:Assessment} found that 59 gradients exceeded the upper bounds of the Conterminous U.S. (CONUS) ionospheric threat model \cite{Datta-Barua10:Ionospheric}, with the largest observed gradient in Brazil of 850.7 mm/km during the local nighttime.
Saito \textit{et al.} \cite{Saito17:Ionospheric} reported the largest gradient of 600 mm/km at nighttime in the Asia-Pacific region.
Therefore, the low-latitude ionospheric threat model that we used in this work is a combination of the CONUS threat model \cite{Datta-Barua10:Ionospheric} for daytime and the 850.7 mm/km (Galeão) or 600 mm/km (Ishigaki and Chennai) worst-case gradient for nighttime, which is the most conservative approach. The mid-latitude ionospheric threat model that we used in this study is the CONUS threat model \cite{Datta-Barua10:Ionospheric}, as it is known to be conservative for mid-latitude regions \cite{Mayer09:Ionosphere, Kim15:GBAS}.

To compare the performance of the optimal $\sigma_{\mathrm{pr}_\mathrm{gnd}}$ inflation method with existing methods, the simulation conditions were set to mimic those reported previously in \cite{Lee11:Ionospheric} and \cite{Seo12:Targeted}. The RTCA 24-satellite GPS constellation was assumed, parameter inflation was performed every minute over a 24-hour simulation period, a minimum $P_k$ value of 0.000180 \cite{Yoon19:Assessment} was used, and the same nominal error models as \cite{Lee11:Ionospheric} and \cite{Seo12:Targeted} were used. However, unlike \cite{Lee11:Ionospheric} and \cite{Seo12:Targeted}, our work focused on low-latitude regions. Hence, a minimum $\sigma_\mathrm{vig}$ of 14.0 mm/km \cite{Chang19:Assessment} and the low-latitude ionospheric threat model were used for low-latitude simulations, instead of 6.4 mm/km \cite{Murphy10:GBAS, Park10:Enabling} and the mid-latitude threat model for mid-latitude simulations.

For MIEV calculations, Lee \textit{et al.} \cite{Lee11:Ionospheric} suggested an expression that considers the magnitude and sign change of the smoothed differential range error following the impact of a severe ionospheric gradient. The ratio of the maximum negative error to the maximum positive error was defined as the ``$c$ factor.'' We considered $c$ factors of 1.0 and 0.5. A $c$ factor of 1.0 indicates that the maximum negative error can be as large as the maximum positive error during range smoothing following the impact of the ionospheric gradient. This is the most conservative assumption. Alternatively, a $c$ factor of 0.5 is a more realistic assumption, which was suggested and used in \cite{Lee11:Ionospheric} and \cite{Seo12:Targeted}.

In the context of precision approaches to a single airport, any significant GNSS jamming \cite{Euro21, Kim22:First} is likely to affect several satellites at once, as opposed to simply increasing the independent outage probability for an individual satellite.  Therefore, jamming is a separate threat, which was not part of our simulation, that, by itself, could eliminate precision approach availability and needs separate mitigation techniques or alternative systems \cite{Park2021919, Rhee21:Enhanced, Lee22:SFOL, Jia21:Ground}.

\subsection{Simulation Results}

\begin{figure}
    \centering
    \includegraphics[width=0.95\linewidth]{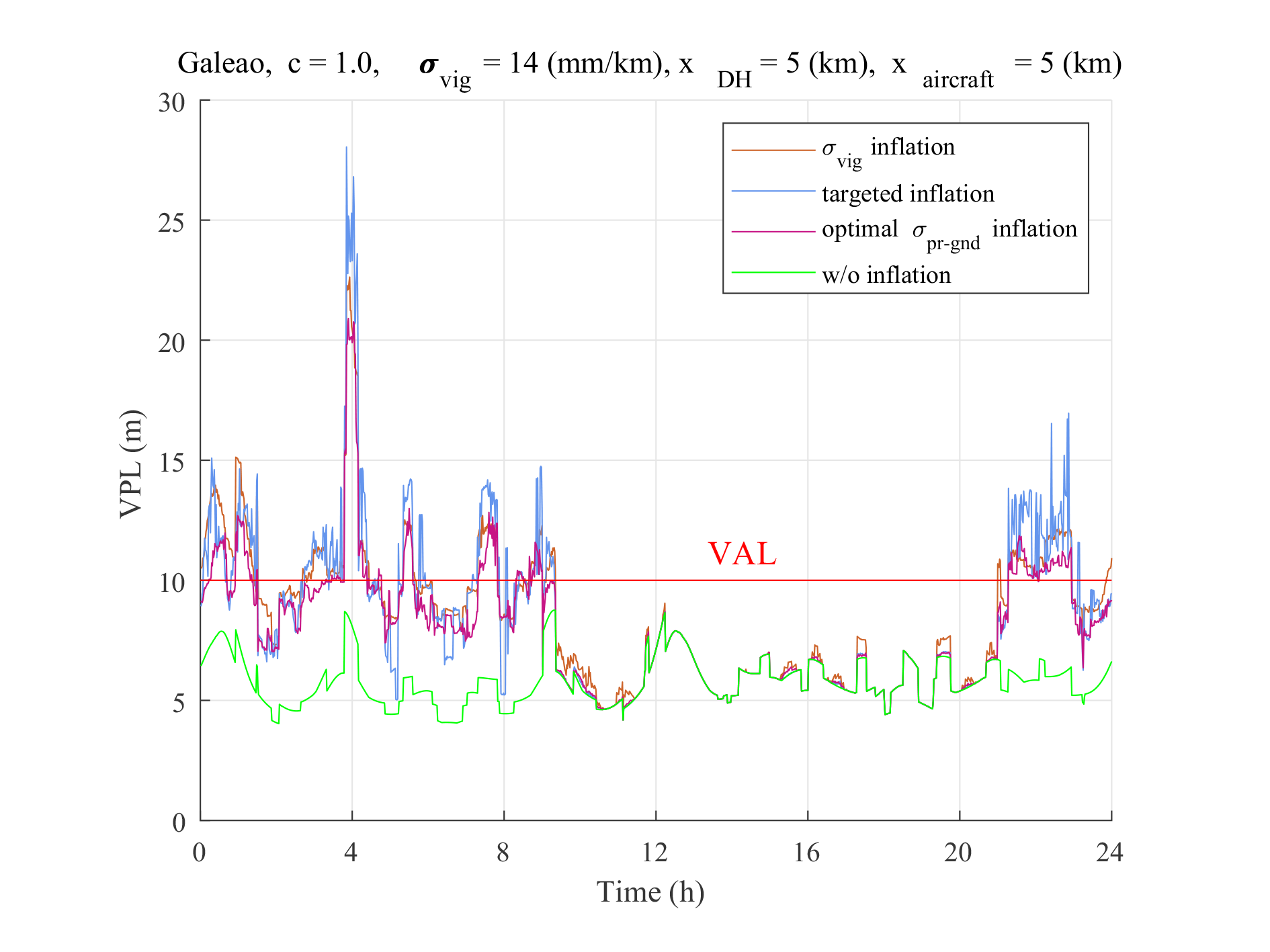}
    \caption{VPL comparison among three geometry-screening algorithms (i.e., $\sigma_\mathrm{vig}$ inflation, targeted inflation, and optimal $\sigma_{\mathrm{pr}_\mathrm{gnd}}$ inflation) at Galeão airport in Brazil. Optimal $\sigma_{\mathrm{pr}_\mathrm{gnd}}$ inflation usually produces lower VPLs than other methods. VPLs are calculated for the all-in-view satellite geometry.}
    \label{fig:VPLCompGaleao}
\end{figure}

Fig.~\ref{fig:VPLCompGaleao} compares the inflated VPLs from the three different geometry-screening methods (i.e., $\sigma_\mathrm{vig}$ inflation, targeted inflation, and optimal $\sigma_{\mathrm{pr}_\mathrm{gnd}}$ inflation) for Galeão airport in Brazil. The inflated VPL after optimal $\sigma_{\mathrm{pr}_\mathrm{gnd}}$ inflation was lower (i.e., performance was better) than in the case of $\sigma_\mathrm{vig}$ inflation or targeted inflation for 82.1\% and 69.9\% of nighttime epochs, respectively.

Table~\ref{tab:Availability} quantitatively compares the GBAS availability, i.e., the time percentage that the inflated VPL for the all-in-view satellite geometry is less than VAL. 
When $c = 1.0$ was used, which is the most conservative approach, the optimal $\sigma_{\mathrm{pr}_\mathrm{gnd}}$ inflation provided approximately 7 percentage point improvement over the existing methods at Galeão airport. At the other two low-latitude airports (i.e., Ishigaki and Chennai), it was difficult to conclude whether the targeted inflation method or optimal $\sigma_{\mathrm{pr}_\mathrm{gnd}}$ inflation method performed better between the two in terms of GBAS availability. However, both methods performed better than the $\sigma_\mathrm{vig}$ inflation method. In terms of computational time, optimal $\sigma_{\mathrm{pr}_\mathrm{gnd}}$ inflation outperformed targeted inflation, as discussed in Section~\ref{sec:OptProbForm}.

\begin{table} 
\centering
\caption{Availability comparison among three geometry-screening algorithms (unit: \%)}
\label{tab:Availability}
\vspace{-4mm}
\begin{center}
{\renewcommand{\arraystretch}{1.4}
 \begin{tabular}[c]{>{\centering\arraybackslash}m{0.7cm}>{\centering\arraybackslash}m{2.0cm} >{\centering\arraybackslash}m{1.0cm}>{\centering\arraybackslash}m{1.0cm}>{\centering\arraybackslash}m{2.0cm} }
 \toprule
 \noalign{\vspace{-2pt}}
    {} & \thead{{\normalfont Airport} \\ {\normalfont (Continent)}}
       & \thead{$\sigma_\mathrm{vig}$ \\ {\normalfont inflation}} 
       & \thead{{\normalfont Targeted} \\ {\normalfont inflation}} 
       & \thead{{\normalfont Optimal} $\sigma_{\mathrm{pr}_\mathrm{gnd}}$ \\ {\normalfont inflation}} \\
    \noalign{\vspace{0pt}}
\hline
\noalign{\vspace{2pt}}
\multirow{5}{*}{$c=1.0$} 
 & Galeão (South America) & 72.57 & 72.92 & 79.79 \\
 & Ishigaki (Asia) & 95.97 & 98.13 & 98.13 \\
 & Chennai (Asia) & 98.54 & 99.24 & 100 \\
 & Memphis (North America) & 100 & 100 & 100 \\
 \noalign{\vspace{1pt}}
 \midrule
 \multirow{5}{*}{$c=0.5$} 
 & Galeão (South America) & 78.47 & 81.60 & 86.67 \\
 & Ishigaki (Asia) & 97.22 & 98.33 & 98.40 \\
 & Chennai (Asia) & 100 & 100 & 100 \\
 & Memphis (North America) & 100 & 100 & 100 \\
 \noalign{\vspace{1pt}}
 \bottomrule
\end{tabular}}
\end{center}
\end{table}

It is noteworthy that the availability benefit of optimal $\sigma_{\mathrm{pr}_\mathrm{gnd}}$ inflation is more prominent when a more severe ionospheric gradient is expected. Galeão airport with the 850.7 mm/km \cite{Yoon19:Assessment} observed ionospheric gradient of Brazil is in a more challenging environment than Ishigaki or Chennai airports with the 600 mm/km \cite{Saito17:Ionospheric} observed gradient of the Asia-Pacific region. 

While all three methods provided 100\% availability for Memphis airport, where a significantly lower ionospheric gradient is expected than in the low-latitude regions, the average VPL inflation of targeted inflation or optimal $\sigma_{\mathrm{pr}_\mathrm{gnd}}$ inflation was significantly less than that of $\sigma_\mathrm{vig}$ inflation. (The average VPL inflation of targeted, optimal $\sigma_{\mathrm{pr}_\mathrm{gnd}}$, or $\sigma_\mathrm{vig}$ inflation was 0.46 m, 0.38 m, or 1.65 m, respectively.)  
Less VPL inflation means a greater availability margin, i.e., margin between the inflated VPL and VAL; thus, targeted inflation or optimal $\sigma_{\mathrm{pr}_\mathrm{gnd}}$ inflation performed better than $\sigma_\mathrm{vig}$ inflation at the mid-latitude airport. In this example, optimal $\sigma_{\mathrm{pr}_\mathrm{gnd}}$ inflation provided a slightly higher availability margin than targeted inflation.

When $c = 0.5$ was used for Galeão airport, which is more realistic assumption than $c = 1.0$, the GBAS availability from all three methods was noticeably larger than when $c = 1.0$ was used. This is because of lower MIEV values, which are the results of a smaller $c$ factor \cite{Lee11:Ionospheric}. 
Overall, the optimal $\sigma_{\mathrm{pr}_\mathrm{gnd}}$ inflation enhanced GBAS availability by 5--8 percentage point compared to the other methods.  When all three methods provided 100\% availability at Chennai and Memphis airports, targeted inflation and optimal $\sigma_{\mathrm{pr}_\mathrm{gnd}}$ inflation provided a greater availability margin than $\sigma_\mathrm{vig}$ inflation, as in the case of $c = 1.0$.

\section{Conclusion} 
\label{sec:Conclusion}

Existing position-domain geometry-screening methods do not provide high availability of single-frequency GBAS in Brazil, where the extremely high ionospheric gradient of 850.7 mm/km \cite{Yoon19:Assessment} has been observed. To improve the availability, we proposed an optimal $\sigma_{\mathrm{pr}_\mathrm{gnd}}$ inflation algorithm and demonstrated that it could provide a 5--8 percentage point GBAS availability enhancement for Galeão airport in Brazil. 
Further, performance of the three geometry-screening algorithms was compared across four airports in both low- and mid-latitude regions. 
At Galeão, optimal $\sigma_{\mathrm{pr}_\mathrm{gnd}}$ inflation provided better availability than the other two methods. The performance of targeted inflation and optimal $\sigma_{\mathrm{pr}_\mathrm{gnd}}$ inflation was similar in terms of availability and availability margin at the other three airports, where the ionospheric gradient was less severe.
However, optimal $\sigma_{\mathrm{pr}_\mathrm{gnd}}$ inflation reduced computational time over targeted inflation. Therefore, the optimal $\sigma_{\mathrm{pr}_\mathrm{gnd}}$ inflation algorithm proposed in this study can contribute to extending the benefits of GBAS for intelligent air transportation to low-latitude regions.

The 86.67\% availability of GBAS at Galeão airport that was attainable with optimal $\sigma_{\mathrm{pr}_\mathrm{gnd}}$ inflation is still less than the 95.97\%--100\% availability at the other three airports in Asia and North America. The extreme ionospheric gradients observed in Brazil were caused by the equatorial plasma bubbles (EPBs) \cite{Yoon19:Assessment}. Therefore, a promising future research direction to improve the GBAS availability in Brazil is to design a more realistic and less conservative ionospheric threat model that considers the characteristics of EPBs. The optimal $\sigma_{\mathrm{pr}_\mathrm{gnd}}$ inflation with a realistic threat model is expected to provide further enhancement of GBAS availability.

\bibliographystyle{IEEEtran}
\bibliography{mybibfile, IUS_publications}

\begin{IEEEbiography}[{\includegraphics[width=1in,height=1.25in,clip,keepaspectratio]{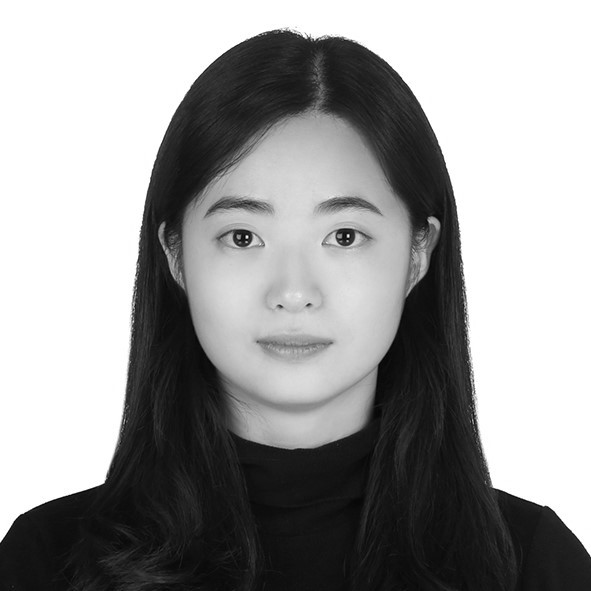}}]{Halim Lee} is an M.S./Ph.D. student in the School of Integrated Technology, Yonsei University, Incheon, Korea. She received the B.S. degree in Integrated Technology from Yonsei University. Her research interests include aviation integrity and opportunistic navigation.
Ms. Lee received the Undergraduate and Graduate Fellowships from the Information and Communications Technology (ICT) Consilience Creative Program supported by the Ministry of Science and ICT, Republic of Korea. 
\end{IEEEbiography}

\begin{IEEEbiography}[{\includegraphics[width=1in,height=1.25in,clip,keepaspectratio]{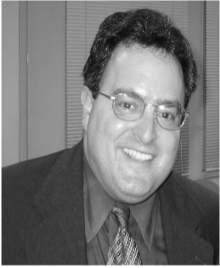}}]{Sam Pullen} is the technical manager of the Ground Based Augmentation System (GBAS) research effort at Stanford University, where he received his Ph.D. in Aeronautics and Astronautics (1996).  He has supported the FAA and other service providers in developing system concepts, technical requirements, integrity algorithms, and performance models for GBAS, SBAS, and other GNSS applications and has published over 100 research papers and articles. He has also provided extensive technical support on GNSS, system optimization, decision analysis, and risk assessment through his consultancy, Sam Pullen Consulting. He was awarded the ION Early Achievement Award in 1999.
\end{IEEEbiography}

\begin{IEEEbiography}[{\includegraphics[width=1in,height=1.25in,clip,keepaspectratio]{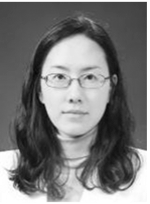}}]{Jiyun Lee} (M'12) received the B.S. degree in astronomy and atmospheric science from Yonsei University in Seoul, Republic of Korea, the M.S. degree in aerospace engineering sciences from the University of Colorado at Boulder, Boulder, CO, USA, and the Ph.D. degrees in aeronautics and astronautics from Stanford University, Stanford, CA, USA, in 2005.
She is an Associate Professor in the Department of Aerospace Engineering at Korea Advanced Institute of Science and Technology in Daejeon, Republic of Korea. As part of her professional experience, she worked as a Consulting Professor with Stanford University, a Principal Systems Engineer with Tetra Tech AMT, and a Senior GPS Systems Engineer with SiRF Technology, Inc. She has published over 80 research papers in the field of GNSS applications, multi-sensor navigation, safety-critical systems, atmospheric science and remote sensing. She was awarded the FAA Recognition Award in 2013.
\end{IEEEbiography}

\begin{IEEEbiography}[{\includegraphics[width=1in,height=1.25in,clip,keepaspectratio]{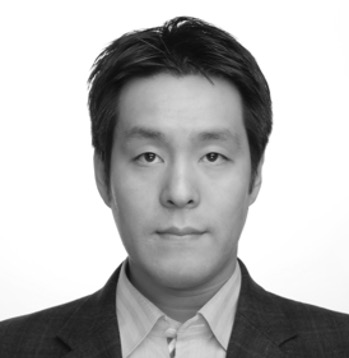}}]{Byungwoon Park} (M'20) received his B.S., M.S., and Ph.D. degrees from Seoul National University, Seoul, Republic of Korea. From 2010 to 2012, he worked as a senior and principal researcher at Spatial Information Research Institute in Korea Cadastral Survey Corp. Since 2012 he has been an Associate Professor at the Aerospace Engineering at Sejong University. His research interests include Wide Area DGNSS (WAD) correction generation algorithms, geodesy and Real Time Kinematics (RTK)/Network RTK related algorithms, and estimation of ionospheric irregularity drift velocity using ROT variation and spaced GNSS receivers
\end{IEEEbiography}

\begin{IEEEbiography}[{\includegraphics[width=1in,height=1.25in,clip,keepaspectratio]{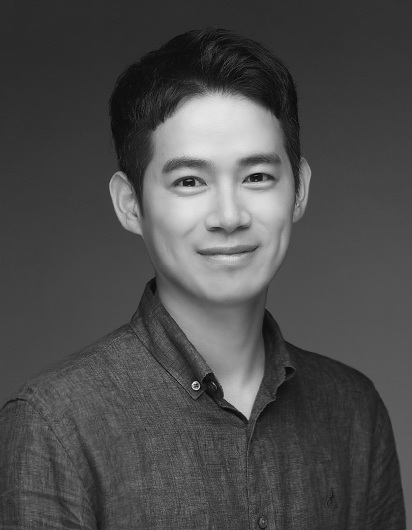}}]{Moonseok Yoon} received a B.S. degree from Korea Aerospace University, Goyang, South Korea, and an M.S. degree and a Ph.D. degree from the Korea Advanced Institute of Science and Technology, Daejeon, South Korea, in 2018, all in aerospace engineering. He is currently a Senior Researcher in the Korea Aerospace Research Institute, Daejeon, South Korea. His current research mainly focuses on atmospheric remote sensing using Global Navigation Satellite System (GNSS) and GNSS-based navigation systems. 
\end{IEEEbiography}

\begin{IEEEbiography}[{\includegraphics[width=1in,height=1.25in,clip,keepaspectratio]{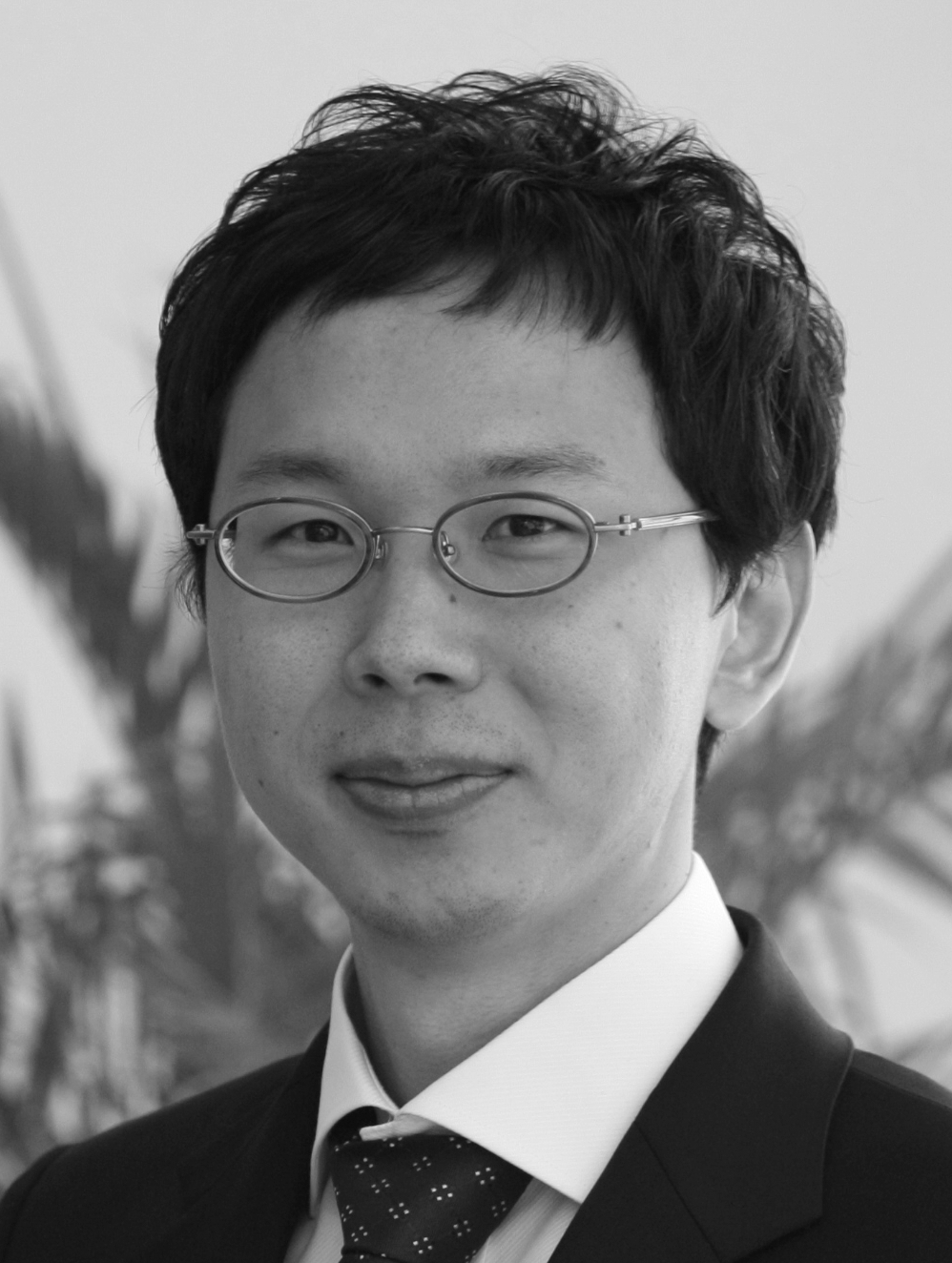}}]{Jiwon Seo} (M'13) received the B.S. degree in mechanical engineering (division of aerospace engineering) in 2002 from Korea Advanced Institute of Science and Technology, Daejeon, Korea, and the M.S. degree in aeronautics and astronautics in 2004, the M.S. degree in electrical engineering in 2008, and the Ph.D. degree in aeronautics and astronautics in 2010 from Stanford University, Stanford, CA, USA. He is currently an associate professor with the School of Integrated Technology, Yonsei University, Incheon, Korea. His research interests include GNSS anti-jamming technologies, complementary PNT systems, and intelligent unmanned systems. Prof. Seo is a member of the International Advisory Council of the Resilient Navigation and Timing Foundation, Alexandria, VA, USA, and a member of several advisory committees of the Ministry of Oceans and Fisheries and the Ministry of Land, Infrastructure and Transport, Korea.
\end{IEEEbiography}

\vfill

\end{document}